\documentclass[aps,floatfix,pre,twocolumn]{revtex4-1}
\usepackage{latexsym,amssymb,graphicx,amsmath,epsfig,calc,times,chemarr,bm}
\usepackage{multirow}

\def\Pv{P_v(N)}

\usepackage[usenames]{color}
\usepackage[ulem=normalem]{changes}

\begin{document}

\newcommand{\change}[1]{\textcolor{Red}{#1}}

\title{Simulations of HIV capsid protein dimerization reveal the effect of chemistry and topography on the mechanism of hydrophobic protein association}
\author{Naiyin Yu}
\author{Michael F. Hagan}
\email{hagan@brandeis.edu}
\affiliation{Martin Fisher School of Physics, Brandeis University, Waltham, MA, 02454, USA.}

\begin{abstract}
Recent work has shown that the hydrophobic protein surfaces in aqueous solution sit near a drying transition. The tendency for these surfaces to expel water from their vicinity leads to self assembly of macromolecular complexes. In this article we show with a realistic model for a biologically pertinent system how this phenomenon appears at the molecular level. We focus on the association of the C-terminal domain (CA-C) of the human immunodeficiency virus (HIV) capsid protein. By combining all-atom simulations with specialized sampling techniques we measure the water density distribution  during the approach of two CA-C proteins as a function of separation and amino acid sequence in the interfacial region. The simulations demonstrate that CA-C protein-protein interactions sit at the edge of a dewetting transition and that this mesoscopic manifestation of the underlying liquid-vapor phase transition can be readily manipulated by biology or protein engineering to significantly affect association behavior. While the wild type protein remains wet until contact, we identify a set of \emph{in silico} mutations, in which three hydrophilic amino acids are replaced with nonpolar residues, that leads to dewetting prior to association. The existence of dewetting depends on the size and relative locations of substituted residues separated by nm length scales, indicating long range cooperativity and a sensitivity to surface topography. These observations identify important details which are missing from descriptions of protein association based on buried hydrophobic surface area.
\end{abstract}

\date{\today}
\maketitle

\section{Introduction}
The hydrophobic effect provides a crucial driving force for the self-assembly of proteins into many biological complexes, such as viral protein coats or capsids \citep{Ceres2002,Kegel2004}, cytoskeletal filaments \citep{Vulevic1997}, and amyloid fibrils (e.g. Ref. \citep{Paravastu2008,Sachse2008}). While the surfaces of unassembled proteins are wet in solution \citep{Cheng1998,Zhou2004}, assembly leads to contact surfaces that are dry \citep{Larsen1998,Rodier2005}. It has been recently shown that hydrophobic protein surfaces sit near a drying transition, enabling them to form soft interfaces with water that lead to assembly.  This paper shows how this phenomenom arises in a realistic model.

Many models for self-assembly and other biological association phenomena assume that binding energy is correlated to buried hydrophobic surface area (e.g. \citep{Eisenberg1986,Spolar1994,Horton1992}). While this generalization has been extremely useful, its accuracy is limited  because it does not account for effects such as surface roughness \citep{Mittal2010}, curvature \citep{Hummer1998,Huang2000}, or long range correlations between chemical groups \citep{Acharya2010}. Corrections arising from these effects will be most important for weak protein-protein interactions, which are ubiquitous in biological systems \citep{Nooren2003} and often essential for the formation of biological assemblages (e.g. Ref. \citep{Zlotnick2003,Ceres2002}). By accounting for the molecularity of water, our study  provides critical details missing from the surface area-based calculation which elucidate how the geometric arrangement and sizes of different chemical groups within a hydrophobic surface determine its interaction.

\begin{figure}[ht]%[bt]
\centering
\includegraphics[angle=0,width=0.9\linewidth]{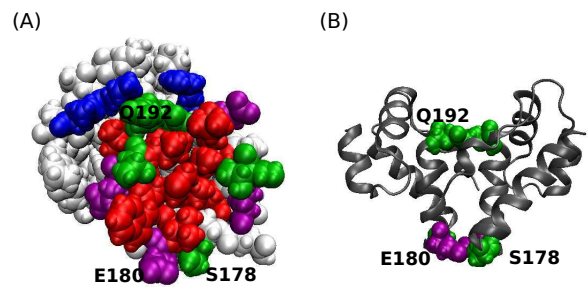}
\caption{The geometry and chemistry of the CA-C dimerization interface. {\bf (A)} A space-filling model of one monomer from the CA-C dimer (PDBID 1A43 \cite{Worthylake1999}). The residues which play a significant role in dimerization \cite{Alamo2003} are color-coded according to residue-type: nonpolar in red, polar in green, basic in blue, and acidic in purple. Three residues which play a key role in determining water behavior, Ser-178, Glu-180, and Gln-192 are labeled. {\bf (B)} A side-view of the dimer interface is shown, with Ser-178, Glu-180, and Gln-192 represented by space-filling models and the remainder of CA-C dimer structure shown in ribbon representation. Images created with VMD \cite{Humphrey1996}.
}
\label{fig:CACstructure}
\end{figure}

 Theoretical work \citep{Stillinger1973,Lum1999,Chandler2005} has shown that hydrophobic association depends on the fact that solvation of a hydrophobic particle exceeding 1 nm in diameter leads to an excess of unsatisfied hydrogen bonds in the surrounding water, which can lead to a state that is close to the liquid-vapor coexistence at ambient conditions. A large, ideal hydrophobic surface (which experiences only repulsive excluded volume interactions with water) pushes the system over a dewetting transition and a liquid-vapor interface is formed \citep{Stillinger1973,Lum1999,Chandler2005}. On the other hand, realistic surfaces such as proteins exert  van der Waals and/or electrostatic interactions that attract the water and thus remain wet. The proximity of an underlying dewetting transition is then only revealed by fluctuations of water density \citep{Patel2010} or the response of water density to perturbations \citep{Huang2000,Chandler2005} such as the confinement introduced by the approach of two such surfaces. If the surfaces are sufficiently close to a dewetting transition, their approach within a critical distance can lead to dewetting and subsequent hydrophobic collapse   \citep{Lee1984,Lum1999,Parker1994,Carambassis1998,Christenson2001,Berne2009,Chandler2005,Liu2005,Giovambattista2008,Krone2008,Patel2012}. However, surfaces of typical proteins found in biological assemblages are geometrically rough and chemically heterogeneous, invariably including hydrophilic groups which locally stabilize liquid water \citep{Hua2009,Acharya2010,Koishi2005,Giovambattista2008,Patel2012}. It is unclear how the principles describing dewetting of idealized surfaces can be applied to more complex protein surfaces.

 Recently Patel and coworkers developed specialized sampling techniques \citep{Patel2010,Patel2011} to measure water density fluctuations in the vicinity of topographically rugged interfaces, and used these to demonstrate that the  model proteins BphC and melittin are close to dewetting transition boundaries \citep{Patel2012}.
Here, we apply these techniques to understand the association of the C-terminal domain of the HIV capsid protein (CA-C), as model system with which to understand the assembly of macromolecular complexes. The size and composition of residues in the interface is typical for protein association interfaces. Furthermore, the CA-C dimerization interface (Fig.~\ref{fig:CACstructure}) plays an essential role in HIV capsid assembly \citep{Mateu2009,Ganser-Pornillos2007,Pornillos2010,Pornillos2009,Pornillos2011,Schwedler2003} and recent studies suggest that it could be a highly effective target for small drug molecules that inhibit assembly \citep{Zhang2009,Bocanegra2011,Dahirel2011}. Isolated CA-C domains dimerize in solution with a dissociation constant of about 10 $\mu$M \citep{Mateu2002} which is similar to that of the full length CA (18 $\mu$M \citep{Gamble1997}), with structures that closely correspond to those found in mature HIV capsids\citep{Pornillos2011}.

We find that, although density fluctuations are enhanced with respect to those found in bulk solution, a dewetting transition does not occur during association of the wild type protein. However, the system is close to a transition as revealed by its sensitivity to perturbations in chemistry or topography. By performing a systematic series of in silico mutations we identify a set of three hydrophilic residues whose simultaneous mutation to nonpolar amino acids of similar size leads to dewetting during association.

 The distances between mutated residues demonstrate cooperative effects on nm length scales, while the dependence of dewetting on the size of the substituting residues indicates that both topography and chemistry control water behavior. These results indicate that the proximate dewetting transition can be manipulated by small perturbations to the CA-C interfacial micro-architecture to effect large changes in association behavior. In contrast to the assumptions underlying traditional surface-area-based estimates of protein association behavior, our results show that the transition to dewetting does not depend only on the total buried hydrophobic surface area or the mean hydrogen bonding potential of the surface, but does depend sensitively on the relative locations and sizes of the mutated residues.

\section*{System}
We perform simulations based on the crystal structure PDBID 1A43 \citep{Worthylake1999}, whose electron density closely fits that found at the hexamer-pentamer interface in electron micrographs of the mature HIV capsid structure \citep{Pornillos2011}.   As shown in Fig.~\ref{fig:CACstructure}, dimerization occurs via the mutual association of $\alpha$-helix 2 (residues S178-V191). The interface involves approximately 1200 \AA$^2$ of buried solvent accessible area contributed by non-polar residues, comprising a hydrophobic `patch' at the center of the contact region which exceeds a nanometer in all directions. The CA-C dimerization interface is thus representative of capsid protein assembly interfaces and other protein-protein association surfaces in terms of structure and composition.
Based on the changes in binding affinity upon mutations of each residue at the dimerization interface, association is primarily driven by hydrophobic interactions but attenuated by electrostatic effects ~\citep{Alamo2003, Alamo2005,Alcaraz2008, Ganser-Pornillos2004,Gamble1997,Yu2009}. In particular, residues whose mutation to alanine significantly impair association are mostly nonpolar, whereas mutation of several polar or charged residues (Ser178, Glu180, Gln192) lead to stronger association \citep{Alamo2005}.

\begin{figure}[h]%[bt]
\centering
\includegraphics[angle=0,width=0.9\linewidth]{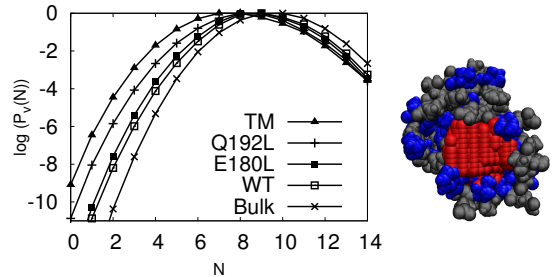}
\caption{The effect of interfacial chemistry on water behavior near isolated CA-C monomers. The distribution of water density fluctuations $P_v(N)$ is shown as a function of the number of waters $N$ in the evaluation volume $v$ (pictured on the right in red) in the vicinity of the CA-C dimerization interface. Distributions are shown for monomers of the wild type protein (WT), and monomers with the indicated sets of mutations: E180L, Q192L and S178A/E180L/Q192L (TM), and a region of bulk water with the same volume as the evaluation region.}
\label{fig:monomer}
\end{figure}

\section{Results}

%\section{Results and Discussion}
%\label{sec:results}
{\bf The CA-C interface on an isolated monomer is wet in solution but leads to enhanced water density fluctuations.} We began by investigating the behavior of water near the dimerization interface of an isolated wild type (WT) CA-C monomer. Because the protein surface has an irregular shape, we employed an extension of the indirect umbrella sampling (INDUS) method \citep{Patel2011, Patel2012} that allows sampling the probability distribution $P_v(N)$ of numbers of water $N$ in an arbitrarily shaped volume $v$ \citep{Patel2012} as described in the Methods. As shown in Fig.~\ref{fig:monomer} the mean number of waters is close to that found in a region of bulk water with the same volume, reflecting the fact that the surface is wet. However, fluctuations to  low densities are enhanced near the interface; i.e. $P_v(N)$ is enhanced at low $N$ in comparison to the distribution for bulk water. This result is consistent with observations on self assembled monolayers and model proteins BphC and melittin \citep{Patel2012}.

To understand how water near the interfacial surface responds to perturbations in the protein sequence, we measured $P_v(N)$ in the same volume for CA-C monomers with hydrophilic amino acids in the interfacial region mutated to hydrophobic amino acids with similar sizes. The measured distributions are shown in Fig.~\ref{fig:monomer} for proteins with one to three such mutations. As shown in Fig.~\ref{fig:monomer}, the single point mutations lead to a small enhancement of low-$N$ fluctuations, and as additional amino acids are mutated the low-$N$ fluctuations are further enhanced. However, the most probable number of waters in the region changes only slightly, and in all cases the surfaces remain wet.

\begin{figure*}[bt]
\centering

\includegraphics[angle=0,width=0.95\textwidth]{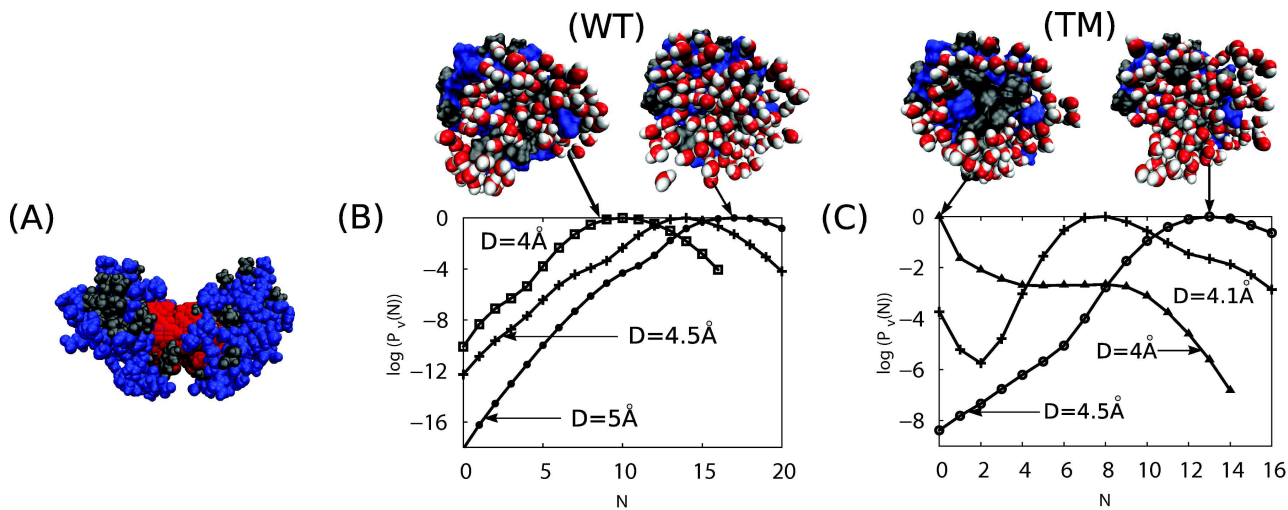}
\caption{Water density distributions during association of CA-C and a mutant. {\bf (A)} The evaluation volume $v$ in red is superposed on the CA-C structure with separation $D=5\,$\AA. {\bf (B)} The distribution of water density fluctuations $P_v(N)$ is shown for wild type (WT) CA-C proteins at separations of $D=4\,$, $4.5\,$ and $5\,$\AA. {\bf (C)} The distribution of water density fluctuations $P_v(N)$ is shown during the approach of two S178A/E180L/Q192L (TM) proteins at separations of $D=4\,$, $4.1\,$ and $4.5\,$\AA. For (B) and (C) representative snapshots are shown in which water within 5 \AA of both proteins is shown along with one monomer oriented to view the interface in cross-section; the solvent accessible surface of the protein is shown, with polar regions colored blue and nonpolar regions colored grey.
}
\label{fig:wildtype}
\end{figure*}

{\bf Water density fluctuations during association.}
To investigate the behavior of water as two CA-C proteins associate, we placed two monomers at different separation distances $D$, and calculated $\Pv$ in the interfacial region between the monomers (Fig.~\ref{fig:wildtype}A). While low-$N$ fluctuations and the probability of drying, $P_v(0)$, increase as the WT monomers approach (Fig.~\ref{fig:wildtype}B), the interface remains wet until the two surfaces come into contact. Specifically, while the most probable value of $N$ decreases with the monomer separation because the interfacial volume decreases, the mean density remains essentially constant and the most probable value of $N$ remains finite. Thus, we conclude that dimerization of wild type CA-C proteins does not involve a dewetting transition.

\begin{figure}[h!]%[bt]
\centering
\includegraphics[angle=0,width=0.7\linewidth]{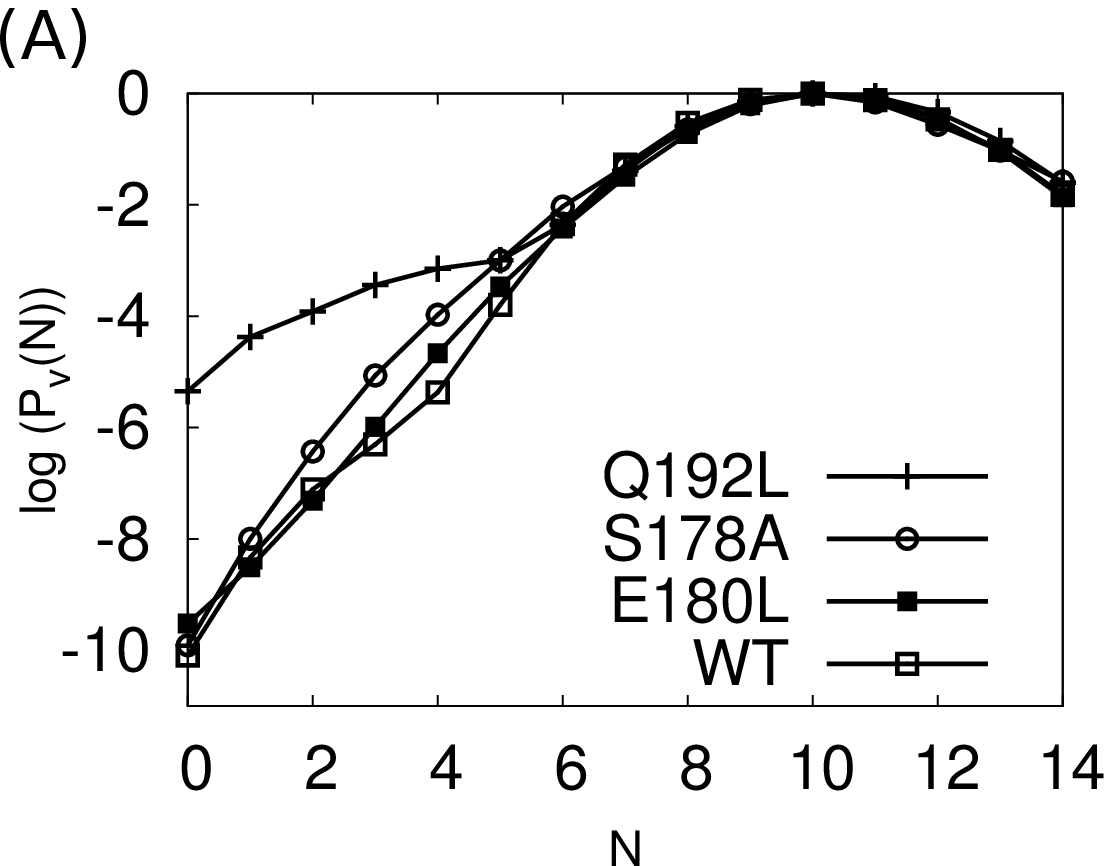}
\includegraphics[angle=0,width=0.7\linewidth]{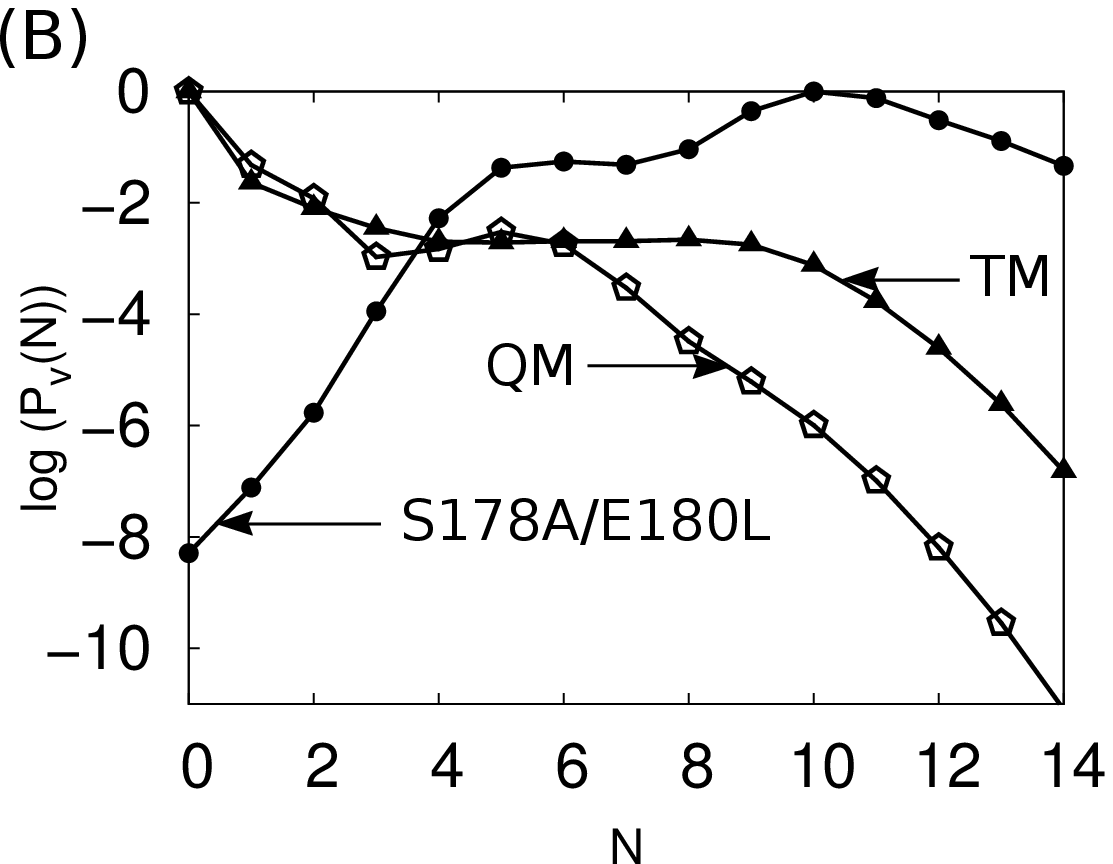}
\caption{Interfacial chemistry determines water behavior during association. The distribution of water density fluctuations $P_v(N)$ is shown for a separation of $D=4$ \AA\ for {\bf (A)} wild type (WT) CA-C proteins and indicated single mutations, and {\bf (B)} the indicated sets of multiple mutations.}
\label{fig:wildtype_mutants}
\end{figure}

We next systematically investigated the effect of the interfacial composition on water behavior during association by measuring $\Pv$ distributions at a separation of $D=4\,$\AA\ for a series of mutations to amino acids in the interfacial region (Fig.~\ref{fig:wildtype_mutants}). We identified no single point or double mutation which led to a dewetting transition, but the mutation Q192L lead to a significant enhancement of low-$N$ fluctuations (Fig.~\ref{fig:wildtype_mutants}A). We also identified a double mutation, E180L/S178A, for which about 5 waters tend to vacate the interfacial region, but dewetting remains relatively improbable. In contrast, the triple mutation S178A/E180L/Q192L (denoted TM) {\bf does} lead to dewetting, with a dramatic change in $\Pv$ and a most probable value of $N=0$ (Fig.~\ref{fig:wildtype_mutants}B). A further mutation S178A/E180L/E187L/Q192L (denoted QM) leads to more aggressive dewetting, as indicated by a higher relative probability of $P_v(0)$.

To identify the critical distance $D_\text{c}$ for dewetting during approach of these mutants, we calculated $\Pv$ for a series of separation distances $D$. As shown in Fig.~\ref{fig:wildtype}C, the surfaces remain wet for distances larger than $D_\text{c}\approx4.1\,$\AA. Near the critical value, the $\Pv$ distributions are bimodal, with high probabilities for the wet (high $N$) and dry (low $N$) states separated by a low probability intermediate region, suggesting the presence of a small barrier to dewetting. A barrier to desolvation could potentially influence the kinetics of dimerization; however, it would be important to determine if such a barrier exists when the proteins associate via different approach vectors.

{\bf  Dependence of dewetting on the location of mutations.} Comparison of $\Pv$ from various protein sequences indicates that mutations are cooperative and that the effect depends sensitively on the relative location of mutated residues. For example, E180L/S178A increases the probability of evacuating 5 waters, but the individual mutations E180L and S178A mutations lead to essentially no enhancement of low-$N$ fluctuations in comparison to the association of wild-type proteins. This observation can be understood by noting that the two monomers are rotated by 180 with respect to each other in the crystal structure, meaning that E180 from monomer A is juxtaposed with S178 in monomer B. Thus, the overlapping hydrophobic area increases only if both amino acids are mutated. On the other hand, the single point mutation Q192L does enhance low-$N$ fluctuations because the Q192 residues from the two monomers partially overlap in the associated structure. Combination of these three mutations (TM) then leads to a sufficient increase in the contiguous overlapping hydrophobic area to give rise to dewetting, even though the mutated amino acids are separated by up to 10 \AA. In contrast, other combinations of 3 mutations which did not lead to as significant changes in overlapping hydrophobic area did not lead to dewetting. Similarly, Q192L/E180L did not enhance low-$N$ fluctuations of the associating monomers relative to those of Q192L. We thus conclude that perturbations which increase the hydrophobic area of an individual monomer have relatively little effect on the propensity for dewetting during association unless the contiguous overlapping hydrophobic area is increased by the mutation.

\begin{figure}[ht]%[bt]
\centering
\includegraphics[angle=0,width=0.7\linewidth]{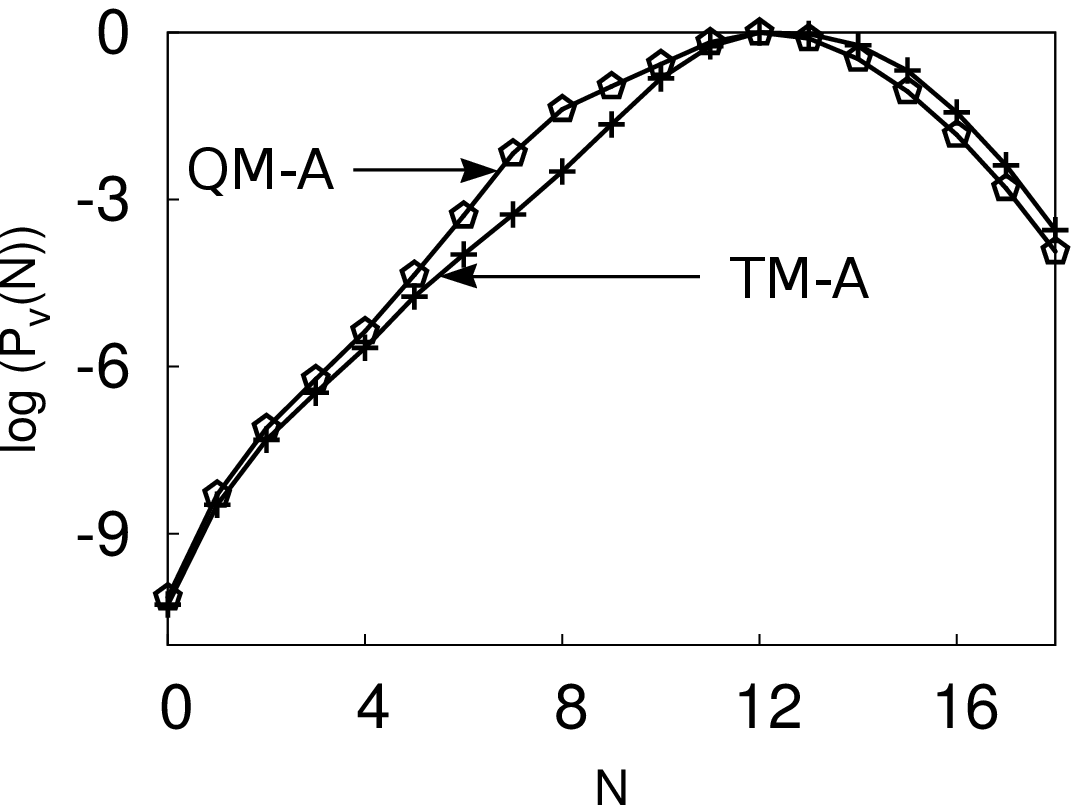}
\caption{The distribution of water density fluctuations $P_v(N)$ is shown for $D=4\,$\AA\ from mutants Q192A/E180A/S178A (TM-A) and Q192A/E180A/S178A/E187A (QM-A). The smaller volume of the substituted ALA residues slightly increases the evaluation volume, leading to a larger mean number of waters than observed for wild type (Fig.~\ref{fig:wildtype_mutants}A).}
\label{fig:ALA}
\end{figure}

{\bf  Dependence of dewetting on the volume of substituted amino acids.} The mutations described in the preceding paragraph were designed to substitute nonpolar amino acids which matched the size and shape of wild type residues as closely as possible. However, the contribution of a particular amino acid or group of amino acids to association is commonly assessed experimentally by substitution to alanine, and the most closely related set of mutations which has been studied experimentally is S178A/E180A/E187A/Q192A (QM-A) \citep{Alamo2005}. Given the sensitivity of dewetting to the location of mutations and other small perturbations, we anticipated that the identity of the residue which is substituted for the wild type amino acid could also affect water behavior and thus we repeated our measurements for triple and quadruple mutants in which all residues were substituted with ALA: S178A/E180A/Q192A (TM-A) and QM-A.  As shown in Fig.~\ref{fig:ALA}, the behavior of these proteins is markedly different from that of TM and QM. Both all-ALA mutation sets exhibit $\Pv$ which are similar to those of wild type, and dewetting is not favorable at any separation distance. This observation can be understood by noting that the substituted alanine residues contribute little nonpolar surface area and their small size in comparison to the wild type residues decreases the topographic complementarity of the associating interfaces.\\

\emph{Hydrogen bonding potential does not imply propensity for dewetting.} It is important to note that the mutations QM and TM do not lead to dewetting simply because they reduce  hydrogen bonding between the dimerization interface and solvent, thereby  reducing the enthalpic barrier to releasing waters during drying. Both the size-preserving mutants (TM and QM) and the all-alanine mutants (TM-A and QM-A) replace the same number of hydrophilic groups at the interface, and thus should lead to an equivalent reduction in the number of hydrogen bonds between the solvated protein interface and water. To verify this supposition, we measured the number of hydrogen bonds among interfacial sidechains and water molecules for the cases of isolated monomers (Mon), the dimer complex (Dim), and monomers separated by $4 \AA$ but solvated (Dim-Sol), for the sequences WT, TM, and TM-A. As shown in Table.~\ref{table:numhbonds}, the number of hydrogen bonds is reduced by the same amount within statistical error for both TM and TM-A as compared to WT. The similarity in hydrogen bonding between TM and TM-A, combined with the observation that TM undergoes a dewetting during association whereas TM-A does not suggests that  hydrogen bonding potential and the propensity for dewetting are not uniquely related. Since the shape of the dimer interface is better perserved in TM proteins, we conclude that the geometric complementarity of hydrophobic regions on association surfaces also plays an important role.

\begin{table}[h]
   \begin{center}
		\begin{tabular}{cccc}
		\hline
		Configuration & Protein & \multicolumn{2}{c}{Number of h-bonds} \\ \cline{3-4}
		& & waters & waters \& sidechains \\		
		\hline
		\multirow{4}{*}{Mon} & WT & 32 & 49 \\
		& TM & 22 & 39 \\
		& TM-A & 22 & 40 \\ \hline
		\multirow{4}{*}{Dim} & WT & 26 & 54 \\
		& TM & 19 & 43 \\
		& TM-A & 21 & 45 \\ \hline
		\multirow{4}{*}{Dim-Sol} & WT & 33 & 52 \\
		& TM & 20 & 40 \\
		& TM-A & 23 & 41 \\ \hline
		\end{tabular}		
    \caption{Number of hydrogen bonds for interfacial residues of the wild type (WT), S178A/E180L/Q192L (TM), and S178A/E180A/Q192A (TM-A) sequences in the case of a solvated monomer (Mon), the dimer complex (Dim), and the two monomer at the verge of association, separated by 4 \AA\ but solvated (Dim-Sol).   The table shows the average number of hydrogen bonds between sidechains of the interfacial residues and (column 3) water or (column 4) water and sidechain atoms. The standard error is approximately 2 hydrogen bonds. The interfacial residues are T148, I150, L151, D152, R154, L172, E175, S178, E180, V181, W184, M185, E187, T188, L189, V191, Q192, N193, K199 and K203.}
    \label{table:numhbonds}
    \end{center}
\end{table}

\section{Discussion and Conclusion}
%\noindent{\bf Discussion.}\\
 Our computational prediction of mutations that alter the CA-C association mechanism could be examined by solution measurement of binding affinities for mutated CA-C proteins or the assembly behavior of mutated full-length protein.  The existence of dewetting implies a stronger hydrophobic contribution to the association free energy as compared to cases without dewetting. Therefore we expect that S178A/E180L/Q192L (TM) or S178A/E80L/E187L/Q192L (QM) should have substantially smaller values of  $K_\text{d}$ than the wild-type CA-C. Notably, comparison of water behavior for the double mutant S178A/E180L and the mutants S178A/E180L/Q192L (TM) or S178A/E180L/E187L/Q192L (QM) (Fig.~\ref{fig:wildtype_mutants}) demonstrates cooperative effects on water behavior of mutations which are separated by more than 10 \AA. This cooperativity arises because the protein interface is already situated at the edge of a dewetting transition. Such a result could be surprising, given that effects of mutations, as measured by the resulting change in an association constant, are usually additive over distances of 6 \AA\ or more \citep{Schreiber1995}.

Such a comparison is complicated by the fact that the mutations will increase the binding affinity by alleviating some repulsions between charged groups in the dimer. E.g., the single point mutations  S178A, E180A, E187A, and Q192A  all led to increase in the dimerization affinity, likely by eliminating electrostatic repulsions in the dimer \citep{Alamo2003,Alamo2005}. This complication could be avoided by comparing dissociation constants for the triple or quadruple mutants with the analogous alanine scanning mutants, TM vs. S178A/E180A/Q192A (TM-A) or QM vs. S178A/E180A/E187A/Q192A (QM-A). Since repulsions between charged groups are eliminated in both cases, we anticipate that the size and shape preserving mutants TM and QM will have increased dimerization affinities in comparison to TM-A and QM-A. Interestingly, the binding affinity has already been measured for QM-A and the results demonstrated long-range nonadditive effects of the mutations, with a smaller increase in dimerization affinity than for Q192A alone \citep{Alamo2005}.

Notably, the amino acids which we have found to most strongly influence dewetting, S178, E180, Q192, are highly conserved among HIV and SIV variants \citep{Kuiken2010} despite that fact they are not necessary for stereospecific binding (see Methods and Ref. \citep{Alamo2005}). Experimental and theoretical investigations have shown that capsid assembly reactions become trapped when protein-protein interactions are too strong  \citep{Ceres2002,Endres2002,Zlotnick2003,Hagan2006,Jack2007,Nguyen2007,Rapaport2010,Hagan2010,Hagan2011}. It is possible that the CA-C interface has evolved to avoid a dewetting transition in order to maintain the relatively weak interactions required for successful assembly.

Since the CA-C  dimerization interface is typical of protein binding surfaces, our results raise the possibility that many proteins sit near a dewetting transition. This scenario is consistent with the idea that biological systems tend to position themselves near phase transitions to enable sensitive regulation \citep{Mora2011}. It can be tested by applying the computational methodologies described here to other proteins to identify mutations which bring about or eliminate dewetting during association, and then experimentally investigating the effects of these mutations on binding affinities. Considering the requirement for weak binding interactions in biological assembly reactions and the prevalence of assembly in biological systems, it is of great importance to extend our understanding of the relationship between sequence, structure, and association behavior to include effects of the molecularity of water, as we have done here for CA-C.

\section*{Methods}
\label{sec:methods}

All simulations were performed using GROMACS 4.0 ~\cite{Hess2008}, modified to enable importance sampling of the distribution of numbers of water molecules within arbitrarily defined regions with the INDUS algorithm \cite{Patel2010, Patel2011}. We used the OPLS-AA force field~\cite{Kaminski2001} to represent protein atoms and the TIP3P model~\cite{Jorgensen1983} to represent water molecules. To check that the choice of force field did not affect the results, additional simulations were performed on the using the Amber 99sb force field \cite{Hornak2006} and SPC/E waters ~\cite{Berendsen1987}.  As described below the results were qualitatively unchanged by the different water model and force field. 

\begin{figure*}[bt]
\centering
\includegraphics[angle=0,width=0.95\textwidth]{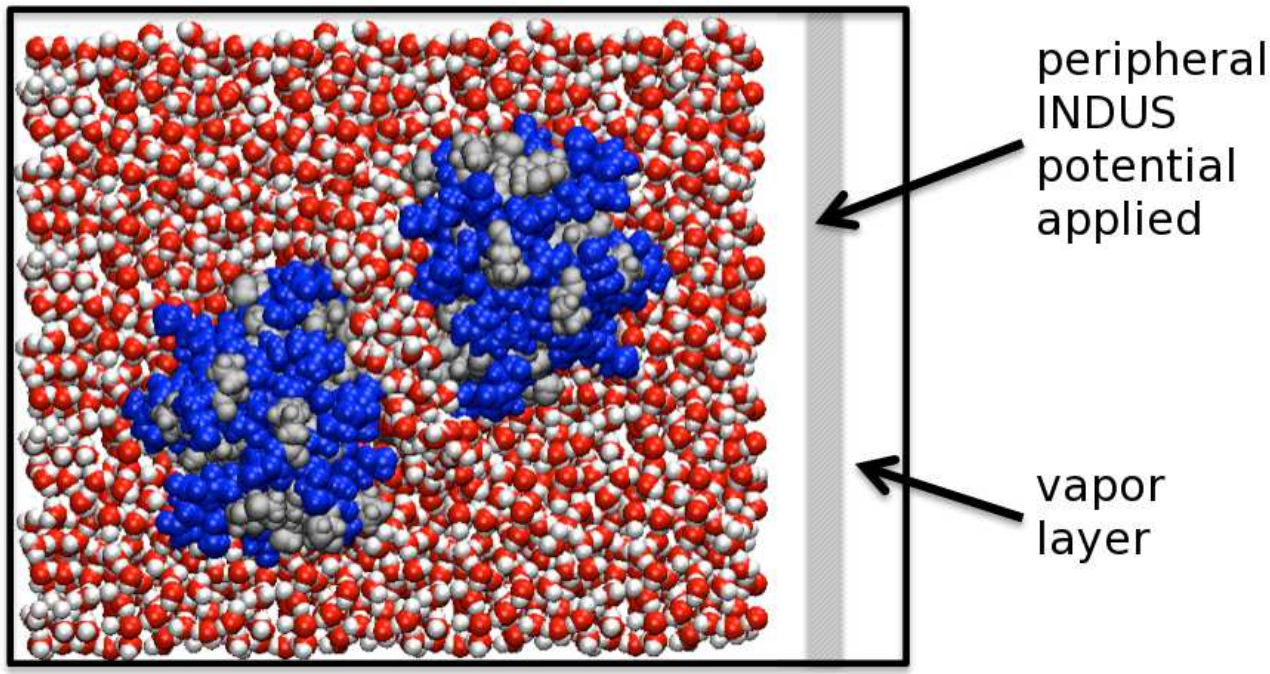}
\caption{Intial set up for water density distribution of HIV capsid protein. The figure shows a cross-section view of the system in a plane orthogonal to the dimerization interface. As shown, the proteins are placed along the diagonal axis of the water box. The region to the right of the water box shows the location of the vapor slab discussed in the text, and the gray bar shows the location of the extra (peripheral) INDUS biasing potential used to ensure that the vapor slab remains at the periphery of the box. The primary INDUS biasing potential applied to the dimer interface region is not shown.}
\label{fig:setup}
\end{figure*}

Our simulations are based on the crystal structure of CA-C, PDB code 1A43 \cite{Worthylake1999}, which corresponds very closely to the structure of two CA-C domains located at hexameric and pentameric interfaces within the mature HIV capsid crystal structure \cite{Pornillos2011}. We prepared a series of initial states in each of which the two monomers were separated by a different distance $D$ along the axis perpendicular to the dimer interface. The protein hydrogen atoms were converted into virtual sites \cite{Bjelkmar2010} to enable a larger timestep. The resulting structure was then solvated in a cuboidal box of water molecules of sufficient dimensions to ensure that protein atoms were separated from box edges by at least $10\,$\AA. This protocol resulted in boxes with $7000\,$  - $8000\,$ water molecules, depending on the separation distance $D$. Sodium and chloride ions were then added to this system to create a bulk salt concentration of 100 mM with additional ions added to achieve charge neutrality. As shown in Fig.~\ref{fig:setup}, the box was then extended by $14\,$\AA\ in the direction perpendicular to the dimer interface to create a vapor slab in order to enable density fluctuations. The position of the vapor slab was held constant at the edge of the box throughout all simulations by applying an extra (peripheral) INDUS potential to bias the center $2\,$\AA\ of the slab toward zero waters. The system then undergoes phase separation, with a region of water and a vapor slab at the periphery. Note that this INDUS potential is in addition to, and separate from, the potential that is applied to bias the number of waters $N$ within the dimer interfacial region toward a specified number. The INDUS potential applied within the vapor slab region has no direct effect on $N$ or its distribution $\Pv$. The presence of the vapor slab enables volume fluctuations of the water region and maintains the system at a pressure equal to the water vapor pressure. The difference between the water vapor pressure and atmospheric pressure has a negligible effect on $\Pv$ since the $PV$ contribution to the free energy of the interfacial region is negligible compared to surface tension contributions for the volumes considered. The distributions $\Pv$ obtained using this method were shown to be equivalent to those obtained from constant pressure simulations using the Langevin piston method in Ref. \cite{Patel2011}. This method of enabling the volume fluctuations was chosen because it does not require calculating the virial contribution from the umbrella biasing forces and because the dynamics of the volume fluctuations may be more realistic than those realized by a more traditional constant pressure algorithm. Simulations were also performed with mutated residues; all mutant structures were built from the wild type by replacing the side chains of the mutated residues using VMD module.

Each system thus generated was minimized when the maximum force is smaller than $500\,$ kJ mol$^{-1}$ nm$^{-1}$ and heated to $T=300$ K. Subsequent molecular dynamics were performed with the backbone atoms of the protein constrained in order to maintain constant relative orientations of the two monomers during the calculation of $\Pv$. Molecular dynamics runs were performed with NVT ensemble and a time step of $4\,$fs. The temperature was kept at $300\,$K using the velocity rescaling modification to the Berendsen thermostat~\cite{Bussi2007}. The protein and solvent were each coupled to separate thermostats with time constants of $0.1\,$ps. The Settle algorithm \cite{Miyamoto1992} was used to constrain H-H and O-H distances in water molecules, and all other bond lengths were constrained using LINCS \cite{Hess1997}. Electrostatic interactions were calculated using the particle-mesh Ewald (PME) algorithm~\cite{Essmann1995}. Van der Waals interactions were switched at $10\,$\AA\ and cutoff at $12\,$\AA. The non-bonded neighbor list is updated every $20\,$fs. Each system was integrated for $10\,$ns ns to ensure equilibration.

After equilibration, the indirect umbrella sampling (INDUS) algorithm ~\cite{Patel2011} was used to calculate distributions of numbers of waters within a `channel' or region between the two dimers for each system. An appropriate definition of the channel is essential to monitor the existence of a dewetting transition; inclusion of additional subregions for which dewetting is unfavorable can obfuscate the results. An initial region was generated by determining the convex hull ~\cite{deBerg2008} defined by the positions of $12\,$ $\alpha$-C residues,  I150, L151, L172, V181, W184, M185, E187, T188, L189, V191, Q192 and N193 at the dimer interface from each monomer. A preliminary INDUS run was then performed to bias the channel toward zero waters by applying a potential of $U_\text{bias}=k n_\text{w}$ with $n_\text{w}$ the number of waters in the channel and $k = 1\,$ kJ nm$^{-2}$. The region was then subdivided into cells of size $1\,$\AA, and cells with an average water occupation greater than $0.01\,$ were eliminated from the channel region. We find that a channel with an hourglass shape naturally arises from this procedure (Fig.~\ref{fig:wildtype}A). We performed the procedure independently for each separation $D$ and set of mutations. Some mutations lead to somewhat larger channel volumes by this procedure; to facilitate comparison of $\Pv$ distributions calculated for defense sets of mutations, the threshold water density within a cell was adjusted to achieve approximately the same channel volume as found for the wild type protein. Once the channel was defined, umbrella sampling was performed in a series of windows,where window centers start from 0 to 15, with interval of 3. For each window, the system was run for 2 ns to allow equilibration under the bias potential followed by 10 ns of data collection. Finally, the unbiased probability distribution was determined using the weighted histogram analysis method (WHAM) \cite{Kumar1992}.

Additional simulations were performed to calculate $P_v(N)$ near isolated monomers. For these runs a single monomer was extracted from the crystal structure, solvated, minimized, and equilibrated as described for the dimer simulations. To facilitate comparison with the dimer simulations, for the wild type monomer and each set of mutations studied, the channel region was defined as the nearest half of the channel determined for the wild type dimer simulation with a separation of $D=4\,$\AA. The distribution $P_v(N)$ within this region was then calculated as described above for the dimer simulations.

Finally, to assess the effect of mutations on the structure of the dimer, we simulated the Q192L/E180L/S178A dimer with a separation $D=0$ for $100\,$ns. We measured an RMSD=$2.6\,$\AA\ between the positions of backbone heavy atoms and those of the wild type crystal structure (for comparison the RMSD=$2.0\,$\AA\ for the wild type structure).  While this is not long enough to definitively determine the stability of the complex, it is consistent with the experimental observation that mutation of the same residues to ALA (Q192A/E180A/S178A) does not impair stability of the complex.

\begin{figure*}[bt]
\centering
\includegraphics[angle=0,width=0.95\textwidth]{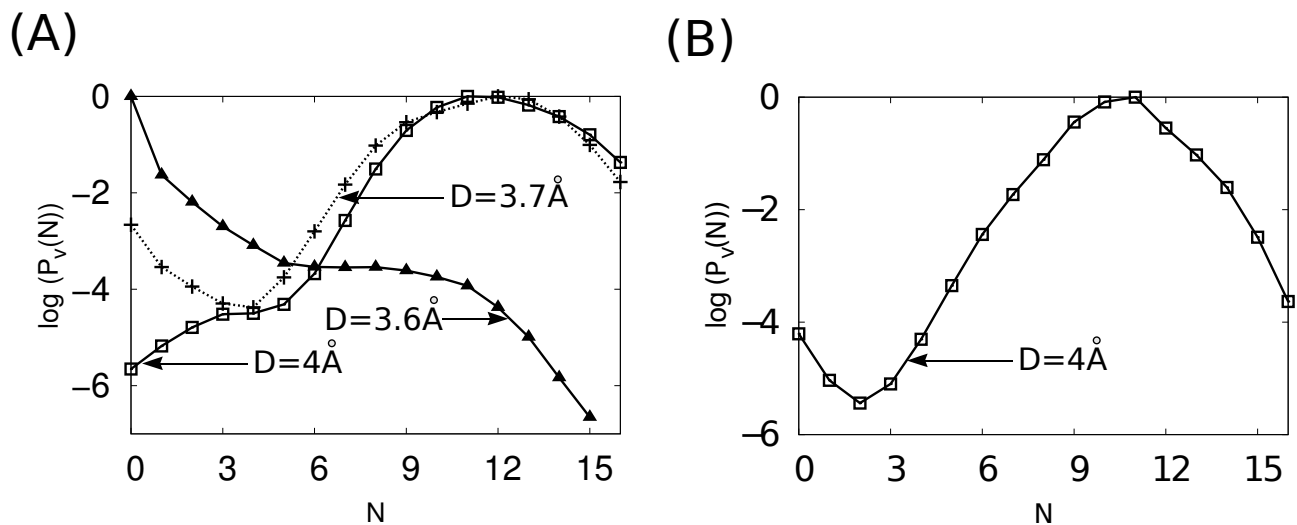}
\caption{Water density distributions calculated using the Amber99sb force field \cite{Hornak2006} and SPC/E waters \cite{Berendsen1987}. {\bf (A)} The distribution of water density fluctuations $P_v(N)$ is shown during the approach of S178A/E180L/Q192L (TM) proteins at separations of $D=3.6\,$, $3.7\,$ and $4\,$\AA. {\bf (B)} The distribution of water density fluctuations $P_v(N)$ is shown for S178A/E180L/E187L/Q192L (QM) CA-C proteins at a separation of $D=4\,$\AA.}
\label{fig:amber}
\end{figure*}

{\bf $P_v(N)$ calculation in Amber force field with SPC/E waters.} Additional simulations were performed on dimerization of the wild type CA-C protein and the mutants TM and QM, using the Amber 99sb force field and SPC/E waters. The behavior of the wild type protein was unchanged with respect to the results described in the main text. As shown in Fig.~\ref{fig:amber}A, the results are qualitatively unchanged for TM and QM proteins, but the critical dewetting distance is shifted to slightly smaller values, approximately 3.7 \AA\, for TM and 4.0 \AA\, for QM. It is not surprising that the critical distance changes slightly upon variations in the water model, since the critical distance is controlled by a balance of  surface tension effects and the chemical potential difference between liquid water and vapor within the interfacial region \cite{Berne2009}.

{\bf Monomer conformational transitions.} Although solution measurement indicates that the CA-C monomer can take an alternative partially unfolded conformation \cite{Mateu2002,Lidon-Moya2005,Alcaraz2007,Alcaraz2008}, thermodynamic and kinetic measurements of the association process are most consistent with a mechanism in which monomers adopt the completely folded conformation prior to association \cite{Lidon-Moya2005}. We did not observe  partial unfolding during our simulations of isolated monomers or monomers at large separation, indicating that such a conformational transition occurs on timescales longer than 100 ns. We thus simulate the approach of two fully folded CA-C monomers.

{\bf Acknowledgements.} 
This work was supported by Award Number R01AI080791 from the National Institute Of Allergy And Infectious Diseases. Computational resources were provided by the National Science Foundation through TeraGrid computing resources (LONI Queenbee and SDSC Trestles) and the Brandeis HPCC. We are most grateful to Patrick Varilly for providing us with a patch to perform INDUS sampling in GROMACS, to David Chandler for comments on manuscript, and to Francesco Pontiggia for insightful discussions.

%\bibliographystyle{unsrt}
%\bibliographystyle{apsrev4-1}
%\bibliography{referencesReformatted}

%Merlin.mbs v4.21 2009-07-09.
%

\end{document}